\documentclass[lettersize,journal]{IEEEtran}
\usepackage{amsmath,amsfonts}
\usepackage{algorithmic}
\usepackage{algorithm}
\usepackage{amsmath,amssymb,amsfonts}
\usepackage{array}
\usepackage[caption=false,font=normalsize,labelfont=sf,textfont=sf]{subfig}
\usepackage{textcomp}
\usepackage{stfloats}
\usepackage{url}
\usepackage{verbatim}
\usepackage{graphicx}
\usepackage{booktabs}
\usepackage{cite}
\usepackage{makecell}
\usepackage{multirow}
\usepackage{bm}
\usepackage[justification=centering]{caption}
\usepackage{adjustbox}
\usepackage[dvipsnames,usenames]{color}
\usepackage{mathrsfs}
\begin{document}
	
	\title{Filter-and-Attend: Wireless Channel Foundation Model with Noise-Plus-Interference Suppression Structure}
	
	\author{
		\IEEEauthorblockN{Yuwei~Wang,~\IEEEmembership{Member,~IEEE,}
			Li~Sun,~\IEEEmembership{Senior Member,~IEEE,}
			Tingting~Yang,~\IEEEmembership{Senior Member,~IEEE,}
			Yuxuan~Shi,~\IEEEmembership{Member,~IEEE,}
			Maged~Elkashlan,~\IEEEmembership{Senior Member,~IEEE,}
			and Xiao Tang,~\IEEEmembership{Member,~IEEE}
		}
		\thanks{
			Yuwei Wang, Li Sun, Tingting Yang, and Yuxuan Shi are with the Department of Broadband Communication, Pengcheng Laboratory, Shenzhen, 518000, China (e-mail: wangyw03@pcl.ac.cn; sunl03@pcl.ac.cn; yangtt@pcl.ac.cn; shiyx01@pcl.ac.cn).}
		\thanks{
			Maged Elkashlan is with the School of Electronic Engineering and Computer Science, Queen Mary University of London, E1 4NS London, U.K. (e-mail: maged.elkashlan@qmul.ac.uk).}
		\thanks{
			Xiao Tang is with the School of Information and Communications Engineering, Xi'an Jiaotong University, Xi'an, 710049, China (e-mail: tangxiao@xjtu.edu.cn).}
	
} 
	
	%
	
		%

	\maketitle
	\thispagestyle{empty}
	\begin{abstract}
		Wireless channel foundation model (WCFM) is a task-agnostic AI model that is pre-trained to learn a universal channel representation for a wide range of communications and sensing tasks. While existing works on WCFM have demonstrated its great potentials in various downstream tasks, the models are all trained using perfect (i.e., error-free and complete) channel information state (CSI) data. In practical systems, however, only degraded CSI obtained from pilot-based channel estimation is accessible, leading to distorted channel representations and performance degradation in downstream tasks for some real-world environments with severe noise and interference. To address this issue, this paper proposes a new paradigm for WCFM, termed as Filter-and-Attend. In this paradigm, Filter refers to explicitly suppressing noise-plus-interference (NPI) in the received signals, while Attend means performing correlation-aware CSI completion and feature extraction using attention mechanism. Specifically, an enhanced WCFM architecture is developed. In this architecture, coarse estimates of the CSIs are first obtained and exploited to construct two projection matrices that extract NPI components in the received signals, which are further processed and removed by a subtraction module. The filtered signal is subsequently passed through a CSI completion network to get a clean CSI for feature extraction. Simulation results demonstrated that compared to the state-of-the-art solutions, WCFM with NPI suppression structure achieves improved performance on various downstream tasks including time-domain channel prediction, frequency-domain channel prediction, and localization.
	\end{abstract}
	
	
	\section{Introduction}
	Future 6G network is featured by the the integration of communications, sensing, and AI. Compared to 2G$\sim$5G networks which are connectivity-oriented, 6G is expected to provide diversified services \cite{You}-\!\!\cite{Saad} (communications, localization, environment reconstruction, edge inference, etc.) across a wide range of scenarios. Moreover, stringent requirements are imposed in terms of the reliability, delay, throughput, and so on.
	
	The fulfillment of the aforementioned 6G visions relies heavily on the availability of accurate knowledge of the underlying wireless channels. However, acquiring channel state information (CSI) in real-time may incur extremely high overhead of pilot signalling in future 6G because of the proliferation of the number of antenna ports, the expansion of frequency bands, and the high mobility of uer equipments (UEs) \cite{YWang-TVT}. Therefore, how to obtain an accurate and universal channel representation with limited pilot overhead emerges as a fundamental and critical challenge in the physical layer design of 6G. Based on existing studies, three types of approaches are identified as promising solutions toward addressing this challenge.
	
	The first type includes traditional model-based CSI measurement methods. To be specific, dedicated reference signals (also known as pilots) are utilized to first estimate the CSIs over some specific resource elements (REs), and then channel interpolation or extrapolation techniques are applied to infer the CSIs of other REs or future instants. While this approach has been widely adopted in current cellular systems, its implementation complexity becomes prohibitively high with the increase in the number of antenna ports. In particular, it is anticipated that at least 256 antenna ports will be supported in 6G systems \cite{Shafi}, which implies that overhead of CSI measurement will be too huge to be affordable. Moreover, the interpolation or extrapolation relies on accurate a priori assumptions of the channel model (e.g., the spatial-temporal-frequency correlation, the noise distribution, and the stationarity), making this approach inherently sensitive to model mismatches.
	
	The second type is based on deep learning techniques \cite{HYe}-\!\!\cite{YWang}. In this paradigm, known pilot signals together with their corresponding received signals are fed into a neural network, which is trained to extract channel features in a data driven manner. This method uses deep neural networks to learn the inherent characteristics of wireless channels, without relying on prior knowledge of channel model, and has demonstrated superior performance in terms of channel estimation and prediction accuracy compared to traditional alternatives \cite{Feifei}-\!\!\cite{Quek}. Nevertheless, the adopted AI models and the extracted channel representations are typically task-specific and scenario-specific, which have limited abilities in generalizing to new tasks or unseen scenarios.
	
	Inspired by the emergence of large language model (LLMs) techniques, the approaches of the third type aim at building foundation models for wireless channel representation to empower multiple physical layer tasks. The foundation models are pre-trained on large-scale wireless channel datasets in a task-agnostic and label-free manner, which imparts them strong generalization abilities across various tasks and scenarios. In \cite{Yucheng}, a LLM-based framework was developed for beam prediction. In this work, the wireless data sequence is reprogrammed into a natural language representation and aligned with the pre-training input format of LLMs, thereby activating the ability of the LLMs to process the wireless time series data. In \cite{Yizhu}, the multi-modal large model DeepSeek was fine-tuned for beam prediction. In particular, positions of the scatters and multi-view images are first fine-tuned with low-rank adaptation (LoRA) to extract environmental embeddings, and then these embeddings are further processed by the large model to output the optimal beam index. In \cite{Liu}, a channel predictor called LLM4CP was proposed to predict downlink channels based on historical uplink channel sequences. This goal is achieved by fine-tuning GPT-2, with customized designs for data preprocessing, embedding, and the output layer. Additionally, the Add-and-Norm layer is fine-tuned as well to match the characteristics of the channel data.
	
	The aforementioned works \cite{Yucheng}-\cite{Liu} rely on the available open-source large language models or multi-modal large models, and utilize fine-tuning techniques to adapt the general-purpose large models to downstream tasks in wireless domain. In contrast with this paradigm, \cite{Boxun}-\cite{Yang} developed wireless native foundation models and performed training from scratch. In \cite{Boxun}, a masked auto-encoder (MAE) based network structure was devised for time-frequency channel prediction tasks, and extensive CSI datasets were used for pre-training such that the model can be generalized to various CSI configurations without any fine-tuning. To further enhance the model's understanding of communications channels, \cite{Feifei2} proposed a decoder-only foundation model architecture CP-GPT, which is trained using both Next Channel Prediction (NCP) loss and Masked Channel Reconstruction (MCR) loss. In \cite{XLiu}, a wireless foundation model was proposed to realize CSI feedback under heterogeneous system configurations. Different from \cite{Boxun}-\cite{XLiu} which address communication tasks, there are also some works studying using foundation models to realize sensing-related tasks such as localization \cite{Aibo}-\cite{Pan}.
	
	The commonality of the papers \cite{Boxun}-\cite{Pan} is that the foundation models developed therein can only handle a single type of task (e.g., channel prediction, CSI feedback, localization, etc.). To enable task generalization, \cite{Sadjad} developed a task-agnostic model called Large Wireless Model (LWM) to generate universal channel embeddings that can be used across a wide range of downstream tasks such as LoS/NLoS classification and beam prediction. In \cite{Yucheng1}, a multi-task prediction model for wireless communication systems was proposed, where a granularity encoding is introduced to distinguish different types of tasks such that the model is able to simultaneously predict CSI, user location, network traffic, etc. In \cite{Yang}, a unified framework of wireless foundation model called WirelessGPT was developed, which integrates cross-domain embedding, positional encoding, and Transformer encoding to generate universal representations of wireless channels. In \cite{Guler}, Guler \emph{et al.} proposed a pre-training method for wireless foundation model that unifies masked reconstruction and contrastive learning to obtain more informative channel representations. \cite{Tianyu} further unleashed the potentials of multi-modal information to integrate the explicit description of the physical environment and the implicit CSI into a universal representation. In this manner, the model well captures multi-level channel features which can be leveraged for various downstream tasks.
	
	While the large model based methods have demonstrated extraordinary performance in physical layer tasks, all of the existing works utilize perfect CSIs across all the REs as the input data to train the model. However, in practical wireless systems where the model is deployed, it is impossible to acquire the perfect CSIs \cite{Sun}. Instead, channel estimation need to be conducted to obtain the estimated CSIs from the known pilots and its associated received counterparts over only a subset of REs. Practical systems typically operate in open environments where signal transmissions are subject to unknown interference and noises, which yield inevitable errors in channel estimation. Based on these erroneous CSIs as inputs, the large AI model will output ``biased" feature representations which do not match the realistic channel status. Consequently, the performance of downstream tasks which highly relies upon these feature representations will be significantly degraded.
	
	To deal with the aforementioned challenge, we propose a novel paradigm for wireless channel foundation model (WCFM) termed as Filter-and-Attend. In this paradigm, the corrupted received signals are first filtered to mitigate noise and interference, after which the refined channel information is exploited by an attention-based CSI completion network and a feature extractor to derive reliable channel representations. Following this paradigm, we design a WCFM architecture with noise-plus-interference (NPI) suppression structure. In our proposed design, coarse channel estimates are first obtained using classical algorithms such as least squares (LS), and then two projection matrices are applied to extract the NPI components within the channel subspace and its orthogonal subspace, respectively. After that, an NPI estimation network is used to estimate the NPI, which will be further subtracted from the received signals. Finally, a channel refinement and completion network is utilized to form a cleaner version of the channel estimate, which is fed into a feature extractor to obtain the channel representation. As will be shown later via simulations, the wireless channel foundation model with the proposed NPI suppression structure achieves improved performance over various downstream tasks including time-domain channel prediction, frequency-domain channel prediction, as well as outdoor localization. The main contributions of this paper are threefold.
	
	\begin{itemize}
		\item First, a new wireless channel foundation model is developed. Compared to all the existing architectures proposed so far, the developed one integrates a NPI suppression structure which facilitates the foundation model to learn and subtract the noise and interference terms from the received signal, thereby producing ``cleaner" CSI estimates to generate high-quality channel feature representations. The NPI suppression module consists of a CSI refinement NN, an NPI estimation and suppression component, and a CSI completion NN.
		\item Second, a CSI completion NN is devised and incorporated into the NPI suppression architecture, which is competent for producing complete CSIs over all REs based on the denoised CSIs across pilot REs. In particular, the CSI estimates on pilots are first encoded into a feature vector using the Transformer network. Then, the time-frequency correlation of CSIs among different REs is extracted based on the feature vector. Given the correlation representation and the time and frequency index of each RE, query vectors are generated, which are used to interpolate the CSI estimates on the pilots to obtain the complete CSI.
		\item Third, a signal-to-interference-plus-noise ratio (SINR) estimator is designed for NPI extraction. In order to realize reliable NPI estimation and cancellation, accurate SINR is required. To achieve this, a 2D Histogram Network is in combined use with a Point Network to well capture the distortion level of the received signal, which can be used to infer the SINR of the system.
	\end{itemize}
	
	The rest of the paper is organized as follows. The system model together with problem formulation is given in Sect. II. Afterwards, the proposed NPI Suppression Module design is presented in Sect. III, where the overall neural network architecture and the details of three building components are shown, and the training method is also elaborated on. Simulation results with comparative studies are exhibited in Sect. IV. Finally, we conclude this paper in Sect. V.
	
	\section{System Model and Problem Formulation}
	In this paper, we consider a MIMO-OFDM communications system with $N$ transmit antennas and $M$ receive antennas. Each slot consists of $T$ OFDM symbols with each containing $K$ subcarriers. Within a slot, the channel matrix over the $k$-th sub-carrier and the $t$-th symbol duration is $\textbf{H}[k,t]\in \mathbb{C}^{M \times N}$, where the entity $h_{ij}$ is the channel coefficient from transmitter antenna \emph{j} to receive antenna \emph{i}. Among the $K\cdot T$ REs over each slot, the set of all RE indices is defined as $\mathcal{A}=\{(k,t)|k=1,...,K,t=1,...,T\}$. Then the channel matrices of all REs are represented by
 \begin{eqnarray}
		\textbf{H}_{\mathcal{A}}=\{\textbf{H}[k,t]|(k,t)\in\mathcal{A}\}\in \mathbb{C}^{KT \times M \times N}.
	\end{eqnarray}
	It is assumed that $K_{p}$ REs (termed as pilot REs hereafter) are occupied by pilots and are used for estimating channels, and the remaining $K_d$ REs (referred to as data REs hereafter) are used for delivering payload data. The index sets of pilot REs and data REs are denoted by $\mathcal{P}$ and $\mathcal{D}$, respectively. The corresponding channel matrices are represented by
	 \begin{eqnarray}
	 	\textbf{H}_{\mathcal{P}}=\{\textbf{H}[k,t]|(k,t)\in\mathcal{P}\}\in \mathbb{C}^{K_p \times M \times N},
	\end{eqnarray} and 
	\begin{eqnarray}
	\textbf{H}_{\mathcal{D}}=\{\textbf{H}[k,t]|(k,t)\in\mathcal{D}\}\in \mathbb{C}^{K_d \times M \times N},
	\end{eqnarray}respectively. For notational convenience, the same subscripts (i.e., $\mathcal{A}$, $\mathcal{P}$, and $\mathcal{D}$) are applied throughout the paper to other RE-dependent quantities, such as symbols, noise, interference, and other channel representations. The transmission process over each RE is expressed as 
	\begin{eqnarray}
		\textbf{y}[k,t]=\textbf{H}[k,t]\textbf{x}[k,t]+\textbf{n}[k,t]+\textbf{i}[k,t],
		\label{trans}
	\end{eqnarray}
	where $\textbf{x}[k,t]\in\mathbb{C}^{N\times1}$ and $\textbf{y}[k,t]\in\mathbb{C}^{M\times1}$ represent the transmitted signal and the received signal, respectively. $\textbf{n}[k,t]\in\mathbb{C}^{M\times 1}$ is the receiver noise, whose elements are independent and identically distributed (i.i.d.) complex Gaussian random variables with mean zero and variance $\sigma^2$. The received interference signal $\textbf{i}[k,t]\in\mathbb{C}^{M\times 1}$ is modeled as 
	\begin{eqnarray}
		\textbf{i}[k,t]=b[k,t]\textbf{G}[k,t]\textbf{z}[k,t],
		\label{inter}
	\end{eqnarray}
	where $b[k,t]$ is a random variable following the Bernoulli distribution $B(1,p)$ with the success probability of $p$, $\textbf{G}[k,t]$ and $\textbf{z}[k,t]$ are the channel matrix from the interferer to the receiver and the transmitted interference signal, respectively. 
	
	For each OFDM slot, the transmitted and received signal pairs over pilot REs $\{\textbf{x}_\mathcal{P},\textbf{y}_\mathcal{P}\}$, where $\textbf{x}_\mathcal{P}=\{\textbf{x}[k,t]|(k,t)\in\mathcal{P}\}$ and $\textbf{y}_\mathcal{P}=\{\textbf{y}[k,t]|(k,t)\in\mathcal{P}\}$, are fed into the wireless channel foundation model to obtain an informative and universal representation of wireless channel, which can be used to realize various communication and sensing tasks such as channel prediction, signal detection, and localization, etc.
		\begin{figure}[t]
		\begin{center}
			\includegraphics[scale=0.60]{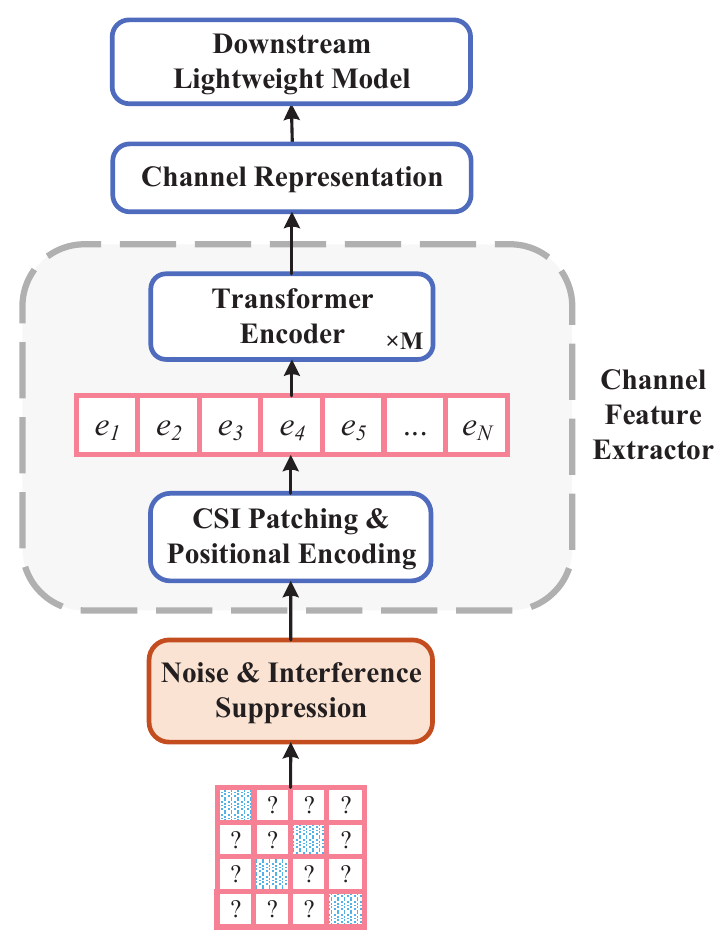}
			\caption{\small{Wireless foundation model with noise-plus-interference suppression structure.}}
		\end{center}
		\label{fig1}
		\vspace{-18pt}
	\end{figure}
	 
	\section{Wireless Foundation Model with NPI Suppression Structure}

	\subsection{Overall Architecture}
	
		\begin{figure*}[t]
		\centering
		\includegraphics[width=7.35in]{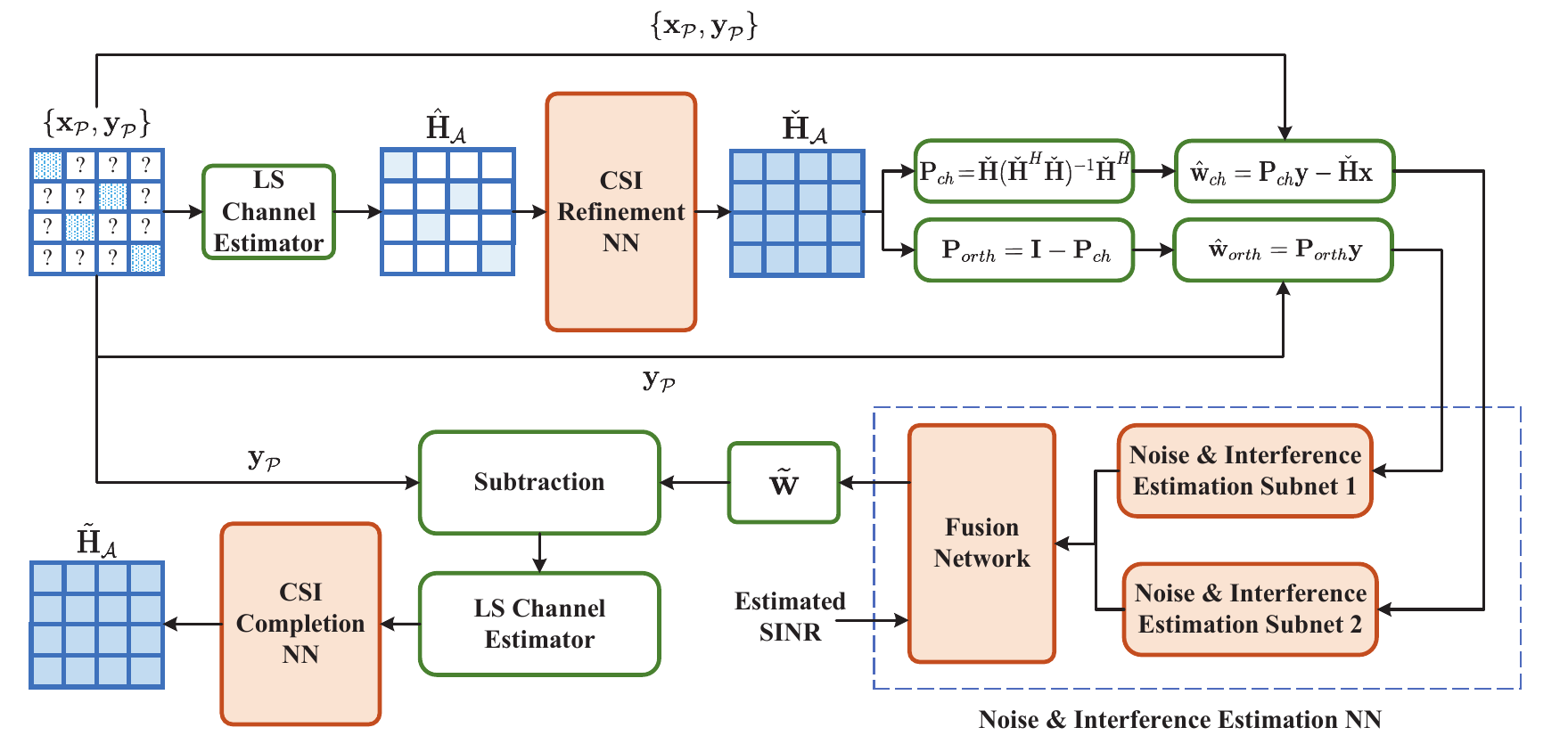}
		\caption{\small{Noise-Plus-Interference Suppression Structure.}}
		\label{fig2}
	\end{figure*}

	As is shown in Fig. 1, we consider a WCFM which adopts a transformer architecture with multi-head attention mechanisms. The backbone of the model follows the common design paradigm in existing wireless foundation models, where CSI is transformed into patches, augmented with the positional encoding, and processed by a stack of transformer encoder blocks to extract representations. Prior works (e.g., \cite{Sadjad}) have demonstrated that transformers are highly effective for modeling wireless channels due to their strong capability in capturing long-range dependencies, spatial–temporal correlations, and structural patterns inherent in CSI.

	However, existing approaches typically assume an access to full and clean CSI over all REs, which is rarely the case in practical systems. In real deployments, only a subset of REs carries pilot symbols, and the received pilot signals are often corrupted by channel noise and interference, which leads to incomplete and degraded CSI, thereby yielding highly-distorted channel representation. To bridge this gap, the proposed architecture integrates an additional noise–plus–interference (NPI) suppression module placed before the transformer backbone. This module restores complete and clean CSI over all REs from the noisy pilot pairs $\{\textbf{x}_{\mathcal{P}},\textbf{y}_{\mathcal{P}}\}$. The NPI suppression module is designed as a plug-in component and can be seamlessly paired with existing transformer-based wireless foundation models without modifying their internal design. Except the NPI suppression module, other modules are almost the same as those adopted by the existing WCFMs in the literature (such as \cite{Sadjad}). Therefore, the following discussion focuses primarily on the design and operation of the proposed NPI suppression module.
	
	The structure of the NPI suppression module is depicted in Fig. 2. Based on known pilots and the associated received counterparts (i.e., $\{\textbf{x}_{\mathcal{P}},\textbf{y}_{\mathcal{P}}\}$), displayed as the dotted blue squares in the upper left part of Fig. 2, a LS algorithm is first applied to form an initial estimate of the CSIs over the pilot REs, denoted by $\hat{\textbf{H}}_{\mathcal{A}}$. Within $\hat{\textbf{H}}_{\mathcal{A}}$, the channel matrix of each REs is 
 	\begin{eqnarray}
\hat{\textbf{H}}[k,t] = 
\begin{cases}
	\dfrac{\textbf{y}[k,t]\textbf{x}^{H}[k,t]}{\textbf{x}^{H}[k,t]\textbf{x}[k,t]}, & (k,t) \in \mathcal{P}, \\
	\textbf{0}, & (k,t) \in \mathcal{D},
\end{cases}
	\label{LS}
\end{eqnarray}
where $(\cdot)^H$ stands for the Hermitian transpose. Afterwards, a CSI refinement neural network (NN) is built to generate the CSIs across all REs $\check{\textbf{H}}_{\mathcal{A}}$. The CSI refinement NN and the CSI completion NN (shown on the lower left part of Fig. 2) have exactly the same structure but different parameters, which will be elaborated on in Section III-B. With supervised training, the CSI refinement NN is able to offer an implicit NPI mitigation capability, thus favoring succeeding processing. Within $\check{\textbf{H}}_{\mathcal{A}}$, the refined CSIs over pilot REs, i.e., $\check{\textbf{H}}[k,t],\forall(k,t)\in\mathcal{P}$, will be used for further NPI suppression. In the rest part of this subsection, since the same operations are applied to the channel matrix of every pilot RE, we omit the RE indices $[k,t]$ for notation simplicity.
	
	Once $\check{\textbf{H}}$ is obtained, we are able to derive two projection matrices $\textbf{P}_{ch}$ and $\textbf{P}_{orth}$, which are expressed as
	\begin{eqnarray}
		\textbf{P}_{ch}=\check{\textbf{H}}(\check{\textbf{H}}^{H}\check{\textbf{H}})^{-1}\check{\textbf{H}}^{H},
		\label{proj-ch}
	\end{eqnarray}
	and
	\begin{eqnarray}
		\textbf{P}_{orth}=\textbf{I}-\check{\textbf{H}}(\check{\textbf{H}}^{H}\check{\textbf{H}})^{-1}\check{\textbf{H}}^{H},
		\label{proj-orth}
	\end{eqnarray}
	respectively, where $\textbf{I}$ is the identity matrix. In ($\ref{proj-ch}$) and ($\ref{proj-orth}$), $\textbf{P}_{ch}$ is the projection matrix onto the channel subspace, and $\textbf{P}_{orth}$ is the projection matrix onto the the subspace that is orthogonal to the channel subspace. For every pilot RE, by multiplying $\textbf{y}$ with $\textbf{P}_{ch}$ and subtracting $\check{\textbf{H}}\textbf{x}$ from the multiplication result, we have
	\begin{eqnarray}
		\hat{\textbf{w}}_{ch}=\textbf{P}_{ch}\textbf{y}-\check{\textbf{H}}\textbf{x},
		\label{w-ch}
	\end{eqnarray}
	which represents the estimated NPI component in the channel subspace. Similarly, multiplying $\textbf{y}$ with $\textbf{P}_{orth}$ yields the estimated NPI component in the subspace that is orthogonal to the channel subspace, given by
	\begin{eqnarray}
		\hat{\textbf{w}}_{orth}=\textbf{P}_{orth}\textbf{y}.
		\label{w-orth}
	\end{eqnarray}
	Due to channel estimation errors, $\hat{\textbf{w}}_{ch}$ and $\hat{\textbf{w}}_{orth}$ contain not only NPI terms, but also residual signal terms. In order to recover the NPI more accurately, a NPI estimation network is introduced, as is shown in the part outlined by the dashed box in Fig. 2.
	
	As illustrated in Fig. 2, $\hat{\textbf{w}}_{ch}$ and $\hat{\textbf{w}}_{orth}$ are passed through two independent NPI estimation sub-networks, whose outputs are then concatenated and fed into a fusion network to form a final estimate of the entire NPI, denoted by $\tilde{\textbf{w}}$. In addition to the concatenated vector, the fusion network also accepts the system SINR as an auxiliary input. The SINR can reflect the accuracy of channel estimation, thereby facilitating the fusion network to identify the amount of residual signal components in the extracted NPI terms. This will help the fusion network accurately extract and combine the actual NPI components from its input. Moreover, since the SINR is related to the power of NPI term in the received signal $\textbf{y}$, it is able to help the fusion network adjust the scale of its output. As a result, the estimated NPI $\tilde{\mathbf{w}}$ can more accurately match the true interference and noise level. In practice, the SINR value is obtained using an SINR estimator, which will be explained in details in Section III-C.
	
		\begin{figure*}[t]
		\begin{center}
			\includegraphics[scale=0.57]{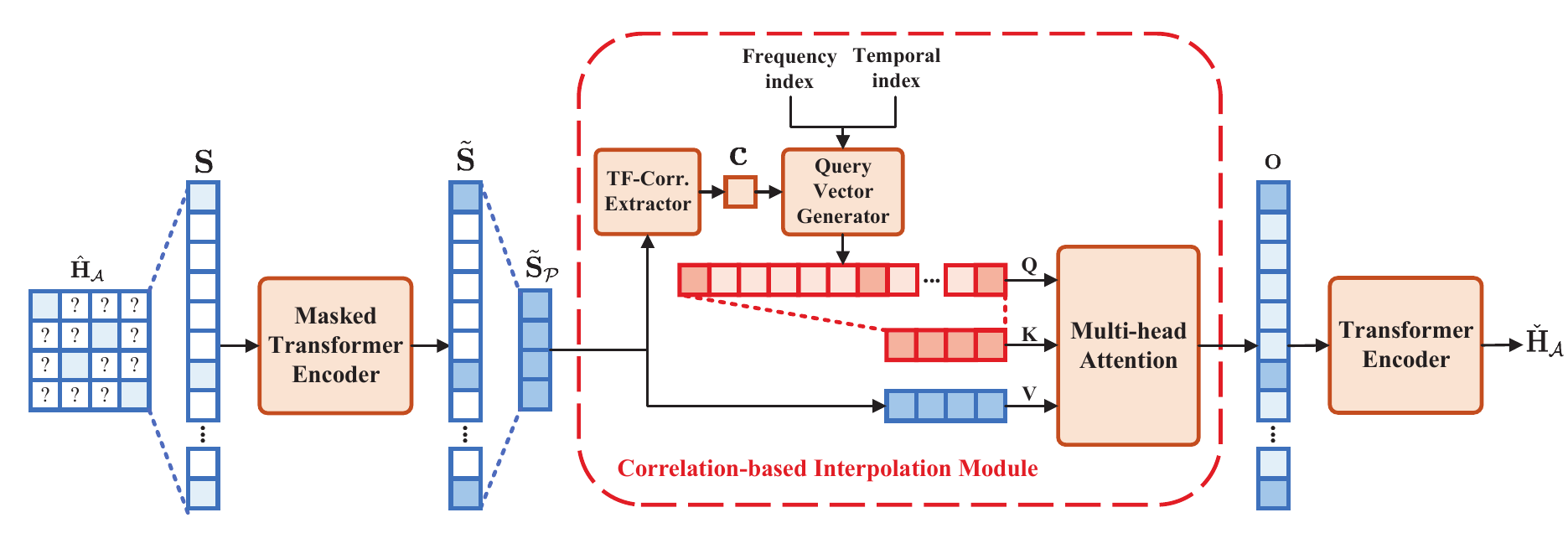}
			\caption{\small{The CSI refinement/completion NN.}}
		\end{center}
		\label{fig3}
	\end{figure*}
	
	Afterwards, $\tilde{\textbf{w}}$ is subtracted from the received pilot symbols $\textbf{y}$, resulting in a cleaner version of the received signal, denoted by $\tilde{\textbf{y}}$. Based on $\tilde{\textbf{y}}$, an LS channel estimator followed by a CSI completion NN is employed to output the refined channel estimates $\tilde{\textbf{H}}_\mathcal{A}$ over all the REs. $\tilde{\textbf{H}}_\mathcal{A}$ will be further processed by the channel feature extractor in the WCFM.
	
	\subsection{CSI Refinement/Completion NN Design}
	
	The CSI refinement/completion NN takes the LS channel estimates over pilot REs $\hat{\textbf{H}}_\mathcal{A}$ as input and outputs the estimated CSIs over all REs $\check{\textbf{H}}_\mathcal{A}$. The structure of the CSI refinement/completion NN is depicted in Fig. 3. $\hat{\textbf{H}}_\mathcal{A}\in \mathbb{C}^{K\times T\times M\times N}$ is first flattened to a real-valued vector sequence $\textbf{S}\in\mathbb{R}^{KT\times 2MN}$, where the sequence length is $KT$. Each element of $\textbf{S}$ is a $2MN$-dimensional real vector formed by concatenating the real and imaginary parts of the channel coefficients corresponding to one RE. Then the vector sequence $\textbf{S}$ is fed into a masked transformer encoder to extract features from the CSI estimates over pilot REs, i.e.,
	\begin{equation}    
		\tilde{\textbf{S}}=f_{\theta_1}(\textbf{S}, \textbf{m})\in \mathbb{R}^{KT\times 2MN},
		\label{MTE}
	\end{equation}
	where $f_{\theta_1}(\cdot)$ denotes the masked transformer encoder, and $\textbf{m}$ is a mask that suppresses the zero vectors corresponding to the data REs. Afterwards, the feature vectors corresponding to the pilot REs within $\tilde{\textbf{S}}$ are taken out to form another vector sequence
	\begin{equation}    
		\tilde{\textbf{S}}_\mathcal{P}=\{\tilde{\textbf{S}}[(k-1)*T+t]|(k,t)\in\mathcal{P}\}\in\mathbb{R}^{K_p\times 2MN}.
	\end{equation}
	Note that instead of directly feeding the pilot-RE vectors only into an unmasked encoder, here we apply a feature extraction with a masked transformer encoder followed by a selection operation. This is to preserve the pilot REs’ positional relationships within the full time–frequency grid. 
	
	The vectors in $\tilde{\textbf{S}}_\mathcal{P}$ containing the channel information of pilot REs are then passed through a correlation-based interpolation module to complete the channel estimates of all REs. The core idea of this module is exploiting the temporal-frequency correlation among different REs to perform interpolation via a multi-head attention layer, in which the query, key, and value matrices are deliberately designed to capture these correlations effectively. In this module, $\tilde{\textbf{S}}_\mathcal{P}$, which contains both the channel information and positional context of the pilot REs, is fed into a temporal-frequency correlation extractor to infer the correlations among all REs, which is represented by a vector $\textbf{c}$. Based on both $\textbf{c}$ and the RE index $(k,t)$, the query vectors of all REs, denoted by $\textbf{Q}\in\mathbb{R}^{KT\times2MN}$, are generated by a fully connected NN. Specifically, the query vector corresponding to the RE at the $k$-th sub-carrier and the $t$-th symbol duration is
	\begin{equation}    
		\textbf{Q}[(k-1)*T+t]=f_{\theta_2}(\textbf{c}, k, t),\quad (k,t)\in\mathcal{A},
	\end{equation}
	where $f_{\theta_2}(\cdot)$ denotes the query vector generator. Then the query vectors associated with the pilot REs are collected to form the key matrix, i.e., 
	\begin{equation}    
		\textbf{K}=\{\textbf{Q}[(k-1)*T+t]|(k,t)\in\mathcal{P}\}\in\mathbb{R}^{K_p\times 2MN},
	\end{equation}
	while $\tilde{\textbf{S}}_\mathcal{P}$  itself serves as the value matrix $\textbf{V}\in\mathbb{R}^{K_p\times 2MN}$. According to the attention mechanism, the scaled inner product between the query vectors and key vectors determines how the value vectors are weighted and aggregated. Therefore, generating both queries and keys based on the correlation descriptor $\textbf{c}$ and the RE indices ensures that the attention weights reflect the underlying temporal–frequency structure of the channel, enabling accurate interpolation from pilot REs to all REs. 
	
			\begin{figure*}[t]
		\begin{center}
			\includegraphics[scale=0.72]{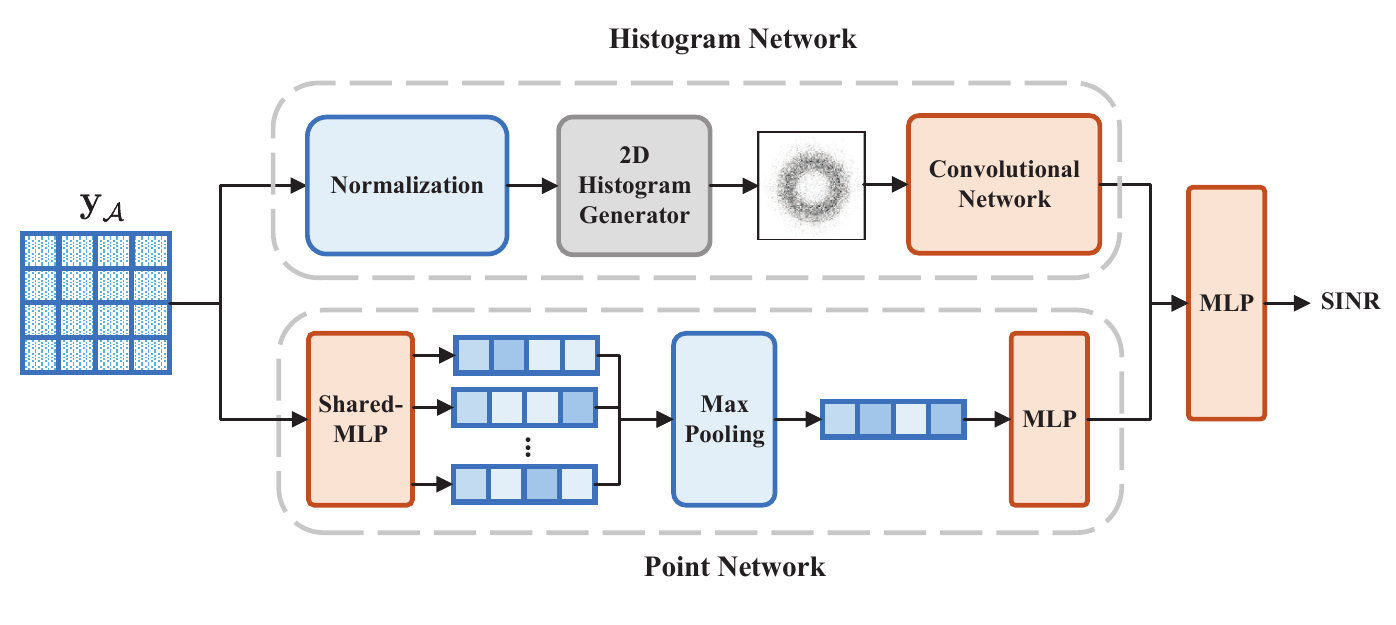}
			\caption{\small{The SINR estimator.}}
		\end{center}
		\label{fig4}
	\end{figure*}

	Then $\textbf{Q}$, $\textbf{K}$, and $\textbf{V}$ are passed through a multi-head attention layer to obtain the features vectors of all REs, denoted by $\textbf{O}\in\mathbb{R}^{KT\times 2MN}$. The attention weights are computed as
	\begin{equation}    
		\textbf{W}=\text{Softmax}(\dfrac{\textbf{Q}\textbf{K}^T}{\sqrt{2MN}})\in\mathbb{R}^{KT\times K_p},
		\label{coeff}
	\end{equation}
	and the output feature vectors are obtained by
	\begin{equation}    
		\textbf{O}=\textbf{W}\textbf{V}.
		\label{weightedSum}
	\end{equation}
	In (\ref{coeff}), the elements of $\textbf{W}$ represent interpolation weights between REs. Consequently, each vector in $\textbf{O}$ is a weighted combination of value vectors, thereby realizing interpolation with the learned weight matrix $\textbf{W}$. Afterwards, $\textbf{O}$ goes through a transformer encoder to produce the channel estimates of all REs $\check{\textbf{H}}_\mathcal{A}$.

	\subsection{SINR Estimator}
	
	The SINR estimator is used to estimate the average SINR over each slot. The estimated SINR allows the NPI estimation network to learn the ``distortion level" of the environment, thereby facilitating the NPI estimation. The structure of the SINR estimator is depicted in Fig. 4, where the received symbols of each slot are used as inputs. The input data is processed by two modules separately: a Histogram Network and a Point Network.
	
			\begin{figure}[t]
		\begin{center}
			\includegraphics[scale=0.45]{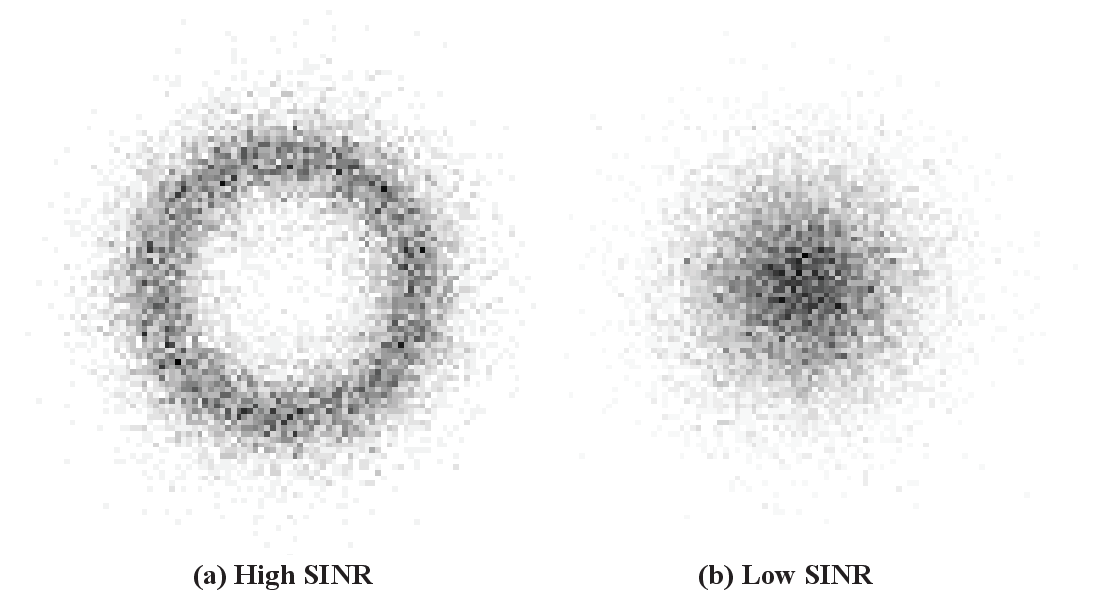}
			\caption{\small{The visualization of 2D histograms for high-SINR and low-SINR cases.}}
		\end{center}
		\label{fig4}
		\vspace{-18pt}
	\end{figure}

	In the Histogram Network, the received symbols $\textbf{y}_\mathcal{A}$ are first normalized to the range of $[-1,1]$ and then fed into a 2D histogram generator to obtain a two-dimensional histogram $\textbf{D}\in\mathbb{R}^{N_D\times N_D}$. The two-dimensional space is divided into $(N_D)^2$ bins, and each element of $\textbf{D}$ is the number of symbol points falling into the corresponding bin. Thus, the 2D histogram shows the statistical distribution of the received I/Q samples within an OFDM slot on the 2D plane. Fig. 5 visualizes the 2D histograms under high-SINR and low-SINR conditions. Each pixel corresponds to a bin in $\textbf{D}$, and the grayscale intensity of the pixel represents the bin value, i.e., the number of received symbols falling into that bin. As is shown in Fig. 5, for QPSK modulated symbols, the histogram pattern varies significantly with SINR. Under high SINR, the distribution of the received symbols is mainly shaped by the channel coefficients across subcarriers, time instances, and receive antennas, exhibiting a clear constellation structure. In contrast, under low SINR, noise and interference dominate, causing the received samples to collapse into a single cluster and blur the constellation. In this sense, the 2D histogram characterizes the amplitude range of received signals, thus reflecting how severe the noise-plus-interference is. The two-dimensional histogram $\textbf{D}$ can be seen as an image with sparse pixels, which is sent into a convolutional neural network to extract the feature for SINR estimation.

	As is mentioned above, the Histogram Network aims at extracting SINR-related information from the overall statistical characteristics of the entire sequence of received symbols. In contrast, the Point Network (PointNet), which is widely used to handle the unordered point cloud data, is employed here to capture fine-grained information from individual received symbols. The use of Point Network for SINR estimation is motivated by the observation that the SINR of an OFDM slot is often dominated by a small subset of informative REs, i.e., REs severely affected by noise or interference. These REs may exhibit distinctive signal characteristics that are not fully captured by statistical representations. 
	
	In the Point Network, the received signal sequence over each RE is first transformed into a $d$-dimensional vector through a shared-MLP layer. With received symbols over $K\cdot T$ REs as the input, we have $K\cdot T$ local feature vectors after the shared-MLP layer. A column-wise max-pooling operation is then applied to aggregate the most informative features across all REs, resulting in a compact global feature representation. Under high-SINR conditions, the received symbols across different REs exhibit weak fluctuations, resulting in similar local features after the shared MLP. In this case, the max-pooling operation produces a stabilized global feature that reflects the common signal-related characteristics shared by most REs. In contrast, under low-SINR conditions, noise and interference introduce significant variability across REs, leading to dispersed local features with pronounced outliers. The max-pooling operation is sensitive to these abnormal REs and highlights their contributions, thereby encoding the strength of noise and interference into the resulting global feature. In summary, the Histogram Network and the Point Network provide complementary perspectives of the received signal. By combining the outputs of the these two networks and passing them through a MLP layer, the SINR value can be estimated.
	
	\subsection{Training Approach}
	
	In the proposed architecture, the CSI refinement network, the NPI estimation network, the CSI completion network, and the SINR estimator are trainable modules, while other parts are non-AI signal processing operations and are thus non-trainable.
	
	The training of these modules experiences three phases. In the first phase, the CSI refinement network and the SINR estimator are trained separately using standard supervised methods with the normalized mean square error (NMSE) as the loss function. In the second phase, the CSI refinement network and the SINR estimator are fixed, and the NPI estimation network is trained following a supervised training method to minimize the NMSE between the estimated and labeled NPI, i.e., 
	\begin{equation}    
		\mathscr{L}_{\text{NPI}}=\text{NMSE}(\tilde{\textbf{w}},\textbf{n}_\mathcal{P}+\textbf{i}_\mathcal{P}),
		\label{eq-loss1}
	\end{equation}
	where $\textbf{n}_\mathcal{P}$ and $\textbf{i}_\mathcal{P}$denote the noise and interference over pilot REs, respectively. After training the NPI estimation network for several rounds, in the third phase, the NPI estimation network will be jointly trained with the CSI completion network to minimize the CSI reconstruction error over all the REs. The loss function is:
	\begin{equation}    
		\mathscr{L}_\text{recon}=\text{NMSE}(\tilde{\textbf{H}}_\mathcal{A},\textbf{H}_\mathcal{A}).
		\label{eq-loss2}
	\end{equation}
	
	\section{Simulation Results and Discussions}
	
		\begin{figure}[t]
		\begin{center}
			\includegraphics[scale=0.3]{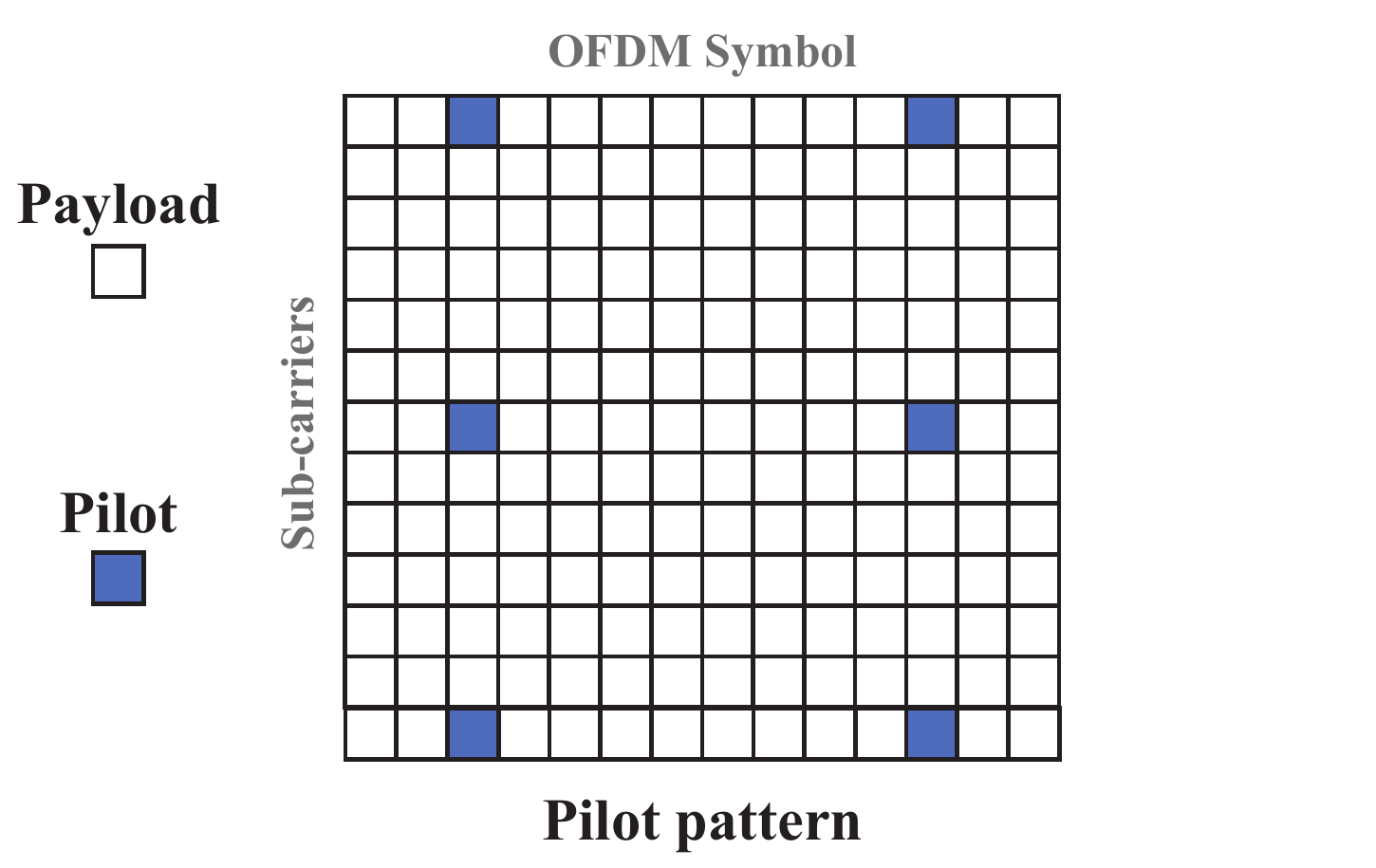}
			\caption{\small{The pilot pattern.}}
		\end{center}
		\label{pilot}
		\vspace{-18pt}
	\end{figure}

	\subsection{Simulation Setups}
	 During the simulations, to evaluate the contribution of the proposed NPI suppression module, the channel feature extractor in Fig. 1 is not trained in our experiments. Instead, we use the NN parameters from the pre-trained LWM \cite{Sadjad} directly. This ensures that the performance gain comes from the NPI suppression module rather than retraining effects. In this section, the open source wireless channel generator DeepMIMO \cite{Alkhateeb} is utilized to produce the CSI dataset. The total number of scenarios is 21, among which 15 scenarios are used for the pre-training of NPI suppression module, while the remaining 6 scenarios are reserved for downstream task evaluation. For each scenario, 2000 BS-UE CSI samples are collected. Specifically, the data from the \emph{ASU Campus}, \emph{New York}, \emph{Los Angeles}, \emph{Chicago}, \emph{Houston}, \emph{Phoenix}, \emph{Philadelphia}, \emph{Miami}, \emph{San Diego}, \emph{Dallas}, \emph{Austin}, \emph{Santa Clara}, \emph{Fort Worth}, \emph{Columbus}, and \emph{Charlotte} scenarios is used for pre-training. The data collected from \emph{Outdoor1}, \emph{Indianapolis}, \emph{San Francisco}, \emph{Seattle}, \emph{Denver}, and \emph{Oklahoma} are used for the training and testing of the downstream model, where proportions of $R_t$ and $1-R_t$ are allocated for training and testing, respectively. $R_t$ ranges from 0.05 to 0.8 in our experiments.
	
	\begin{table}[t]
	\begin{center}
		\caption{\label{table:1} Simulation Settings}
		\begin{tabular}{ccc}
			\toprule
			Parameter & Symbol & \begin{tabular}[c]{@{}l@{}}Value or Range\end{tabular} \\
			\midrule
			Number of Pre-training scenarios & None & 15\\
			Number of evaluation scenarios & None & 6\\
			Carrier frequency & None & 3.5GHz \\
			Number of BS antennas & $M$ & 32 \\
			Number of UE antennas & $N$ & 1 \\
			Number of sub-carriers &$K$& 32 \\
			Number of OFDM symbols per slot &$T$& 14 \\
			Sub-carrier spacing & None &  30kHz \\
			UE velocity & None &  [0.5,30]m/s \\
			Downstream training proportion & $R_t$ &  [0.05,0.8] \\
			SINR &None& [-8,8] \\
			\bottomrule
		\end{tabular}
	\end{center}
\vspace{-18pt}
\end{table}

	For each scenario, a BS with 32 antennas is deployed at a fixed location, and single-antenna UEs are distributed randomly within a area. The moving velocities of UEs range from 0.5m/s to 30m/s. Each transmission slot consists of 14 OFDM symbols and every symbol spans across 32 subcarriers. As is illustrated in Fig. 6, the 3rd and 12th OFDM symbols are used for pilot transmission, where the pilots are inserted with a spacing of 6 subcarriers. Other REs within each slot are occupied by payload data. The simulation settings are summarized in Table I. 

	Based on the received signals at the BS, numerous $\{\textbf{x}_{\mathcal{A}},\textbf{y}_{\mathcal{A}}\}$ pairs are collected. During the training phase, these pairs with the known CSI and noise-plus-interference label are used to train the network. During the inference phase, Only the $\{\textbf{x}_{\mathcal{P}},\textbf{y}_{\mathcal{P}}\}$ pairs on the pilot REs and the received payload data $\{\textbf{y}_{\mathcal{D}}\}$'s are available at the BS, which are fed into the NPI suppression module and the succeeding channel feature extractor to extract channel representations.
	

	\subsection{Benchmark Schemes and Downstream Tasks}

\begin{figure*}[b]
	\begin{center}
		\includegraphics[scale=0.4]{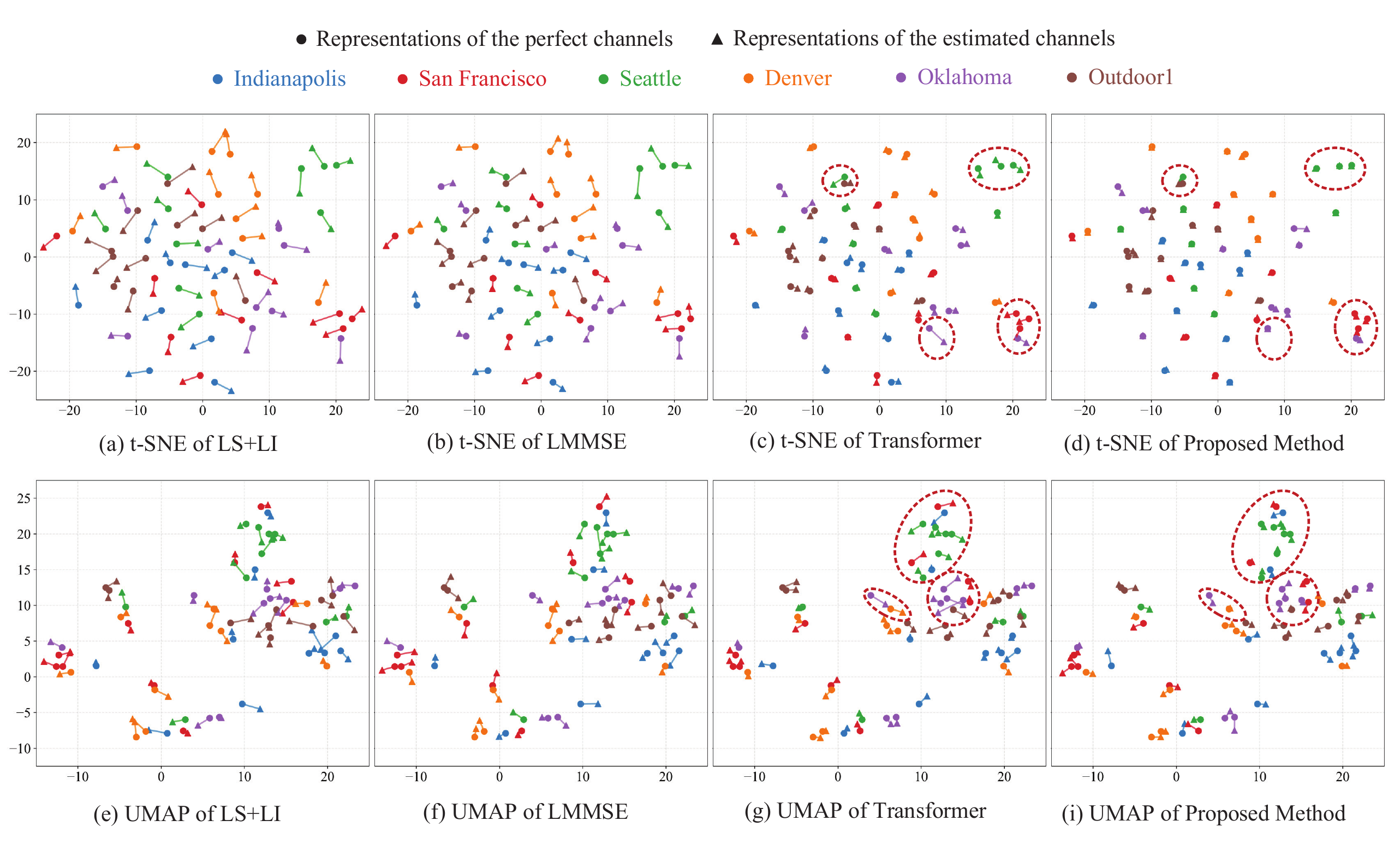}
		\caption{\small{Channel representation visualization of various methods based on t-SNE and UMAP.}}
	\end{center}
\end{figure*}

	For comparison, three competing solutions are chosen as benchmarks in the simulation. For all the candidates, the inputs are the same, i.e., the $\{\textbf{x}_{\mathcal{P}},\textbf{y}_{\mathcal{P}}\}$ pairs on the pilot REs and the received payload data $\{\textbf{y}_{\mathcal{D}}\}$'s, and the outputs are the recovered CSIs over all the REs within each slot. These outputs will be fed to the channel feature extractor to output a universal representation of CSIs, which is further processed by a lightweight ResNet \cite{He} model to obtain the final downstream task results. The benchmarks are briefly introduced as follows.
	
	1) LS plus linear interpolation (LS+LI) \cite{Coleri}: In this approach, a LS channel estimator is fisted applied to estimate the CSI over the REs that carry on pilots, and then a linear interpolation algorithm is utilized to obtain the full CSIs across all REs.
	
	2) LMMSE \cite{Bj}: A LMMSE estimator is used to recover the CSIs of all the REs. This is a widely adopted approach in 5G. Note that the channel covariance matrix is required to be available for this method. In our simulations, an exponential function is adopted to model the time-frequency correlation, from which the channel covariance matrix is derived.
	
	3) Transformer-based method \cite{FLiu}: In this method, the LS algorithm is first used for channel estimation to obtain the CSIs over pilot REs, and then a linear interpolation algorithm along with a Transformer-based network is deployed to recover the full CSI of all REs.
	
	In order to separately evaluate the effectiveness of the proposed CSI refinement NN and the succeeding NPI reduction process, we examine the performance at two different stages: the output of the CSI refinement network $\check{\textbf{H}}_\mathcal{A}$, referred to as ``Proposed Method (Stage 1)'', and the final output of the complete framework $\tilde{\textbf{H}}_\mathcal{A}$, termed as ``Proposed Method (Full)".

	We consider three downstream tasks, introduced as follows.
	
	1) Time-domain channel prediction: This task aims to predict the CSI of the next $N_t$ OFDM symbols given the channel representations of the historical CSI over 14 OFDM symbols. In this task, NMSE between the predicted CSI and the ground truth CSI is used as the evaluation metric. 
	
	2) Frequency-domain channel prediction: This task focuses on predicting the CSI of $N_f$ adjacent subcarriers based on the channel representations of the existing CSI over 32 subcarriers. This task also uses NMSE as the evaluation metric.
	
	3) Outdoor localization: In this task, the location of the UE is estimated based on the channel representations of the CSI between the BS and the UE. In this task, the mean Euclidean error (MEE) between the estimated location and the real location is served as the evaluation metric.

	\subsection{Visualizations of Channel Representations}
	To intuitively demonstrate the effectiveness of different methods in extracting accurate channel representations, the high-dimensional channel representations are projected onto a two-dimensional space using t-distributed stochastic neighbor embedding (t-SNE) \cite{Maaten} and uniform manifold approximation and projection (UMAP) \cite{McInnes}. The t-SNE method emphasizes the preservation of pairwise similarities among neighboring samples, thereby enabling an intuitive visualization of local clustering structures. In contrast, UMAP preserves not only local neighborhood relationships but also the global topological structure of the underlying data manifold. By leveraging these two dimension reduction techniques, the distance between channel representations derived from perfect CSIs (i.e., when the input of the channel feature extractor in Fig. 1 is the clean and complete CSI) and those obtained from estimated CSIs can be directly visualized in the 2D space, as illustrated in Fig. 7. In Fig. 7, dots and triangles denote the representations based on perfect CSIs and estimated CSIs, respectively, and different colors correspond to different scenarios.
	
	As illustrated in Fig. 7(a), (b), (e), and (f), substantial gaps can be observed between the channel representations derived from perfect CSIs and those obtained from the channel estimates using traditional methods (i.e., LS+LI and LMMSE). This indicates that the channel representations extracted from these estimated CSIs are severely distorted by noise and interference. As a result, the errors introduced during the channel estimation stage are directly propagated to the representation extraction process, leading to a significant mismatch with the representations based on perfect CSIs. As shown in Fig. 7(c) and (g), the transformer-based approach is able to effectively mitigate such representation distortion by leveraging its strong modeling capability to capture correlations among REs. 
	
	Nevertheless, for a portion of channel samples, the resulting representations using the Transformer-based approach still exhibit noticeable deviations from the ground truth, especially for those within the red circles. This suggests that although the transformer-based method reduces the overall estimation error, it remains sensitive to noise and interpolation inaccuracies in challenging channel conditions. By further incorporating the proposed noise-plus-interference suppression mechanism together with the CSI completion NN, these highly deviated channel representations are pulled closer to their perfect-CSI counterparts, as shown in Fig. 7(d) and (i). This improvement can be attributed to the joint suppression of NPI during channel estimation and the correlation-aware interpolation performed within the attention mechanism, which together enable the extraction of cleaner and more informative channel representations.
	
	\subsection{Simulation Results - Time-domain Channel Prediction}
	
	\begin{figure}[t]
		\begin{center}
			\includegraphics[scale=0.57]{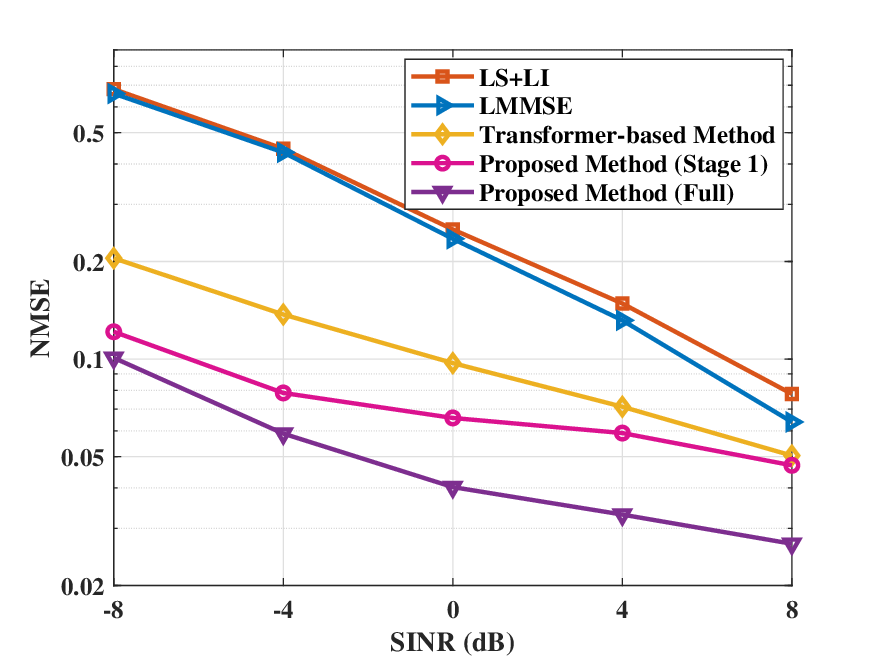}
			\caption{\small{NMSE versus SINR (dB) for time-domain channel prediction task.}}\vspace{-18pt}
		\end{center}
		\label{fig6}
	\end{figure}
	
	The performance of time-domain channel prediction for different methods is presented in Fig. 8, where the number of OFDM symbols to be predicted $N_t$ is 14, and the percentage of downstream training samples $R_t$ is 50$\%$. As expected, the NMSE of all candidates decreases with the increase in SINR. This is because higher SINR leads to more accurate channel estimation, which in turn yields more precise channel representations and thereby improves the performance of the downstream channel prediction. However, the traditional model-based methods (i.e., LS+LI, LMMSE) behave poorly across entire SINR range, since the preset model fails to accurately match various channel conditions. In contrast, the transformer-based method is able to achieve substantial performance gain over model-based methods by effectively extracting temporal-frequency-spatial correlations, which are critical in CSI completion. 
	
	Nevertheless, a considerable performance gap can be observed between the transformer-based method and the proposed method (stage 1), especially in the low-SINR regime. This performance improvement primarily stems from the different interpolation mechanisms adopted by these two methods. Specifically, the transformer-based approach first applies linear interpolation to the LS channel estimates and then refines the interpolated results using a transformer encoder. Consequently, the interpolation errors introduced in the initial linear interpolation stage are directly propagated to the subsequent transformer processing. In contrast, the proposed CSI refinement method avoids such noise propagation by performing correlation-aware interpolation directly using the attention mechanism. By adaptively weighting pilot REs according to learned time–frequency correlations, the proposed method is able to effectively suppresses noise-dominated pilots, leading to significantly improved performance under low-SINR conditions.
	
	As is shown in Fig. 8, by integrating noise-plus-interference suppression mechanism into channel estimation and completion, the proposed method (full) is able to achieve additional performance gain compared with proposed method (stage 1), and this gain becomes more apparent as the SINR increases. At higher SINR levels, the CSI refinement network is able to produce more accurate channel estimates, which in turn enables more reliable estimation of the projection matrices and NPI terms. As a result, the subsequent NPI suppression becomes more effective, leading to increasingly larger performance improvements in the high-SINR regime.

\begin{figure}[t]
	\begin{center}
		\includegraphics[scale=0.57]{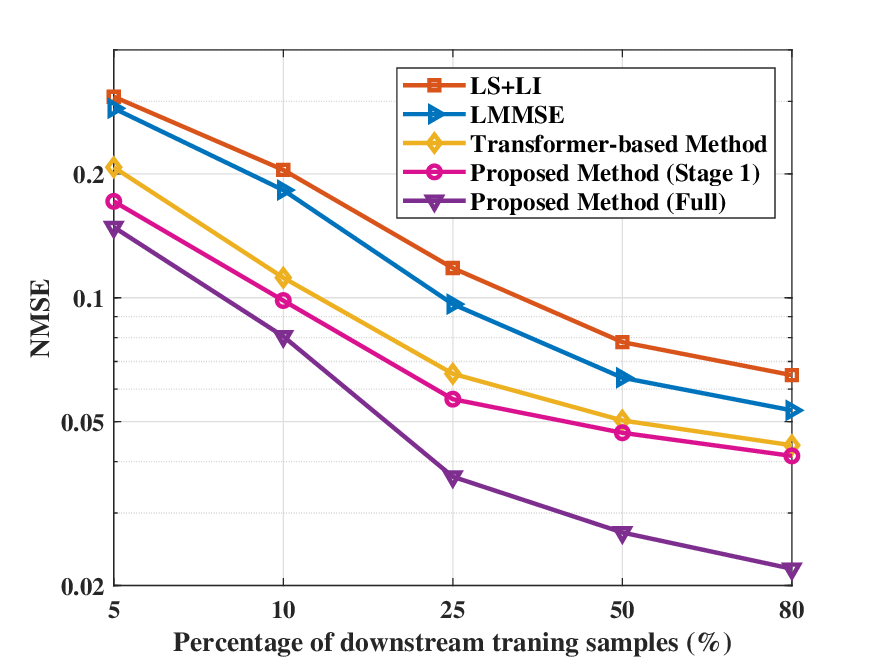}
		\caption{\small{NMSE versus percentage of downstream \\training samples ($R_t$) for time-domain channel prediction.}}\vspace{-18pt}
	\end{center}
\end{figure}

	Fig. 9 illustrates the NMSE trends of various methods with respect to the percentage of downstream training samples ($R_t$), where the SINR is 8dB. As is shown in Fig. 9, the proposed method consistently achieves the lowest channel prediction NMSE across all the considered values of $R_t$, demonstrating its adaptability. Notably, we can observe from Fig. 9 that the proposed method (full) with $R_t=25\%$ is able to outperform all the other solutions with $R_t=80\%$, exhibiting its great potential in mitigating the dependence on large amount of labeled training samples. Thanks to the NPI suppression mechanism, the channel representations produced by the proposed method contains cleaner, richer and more discriminative channel characteristics, which facilitates easier and more data-efficient downstream training and mitigates the over-fitting risk with limited labeled samples.
	
	\subsection{Simulation Results - Frequency-domain Channel Prediction}

	\begin{figure}[t]
		\begin{center}
			\includegraphics[scale=0.57]{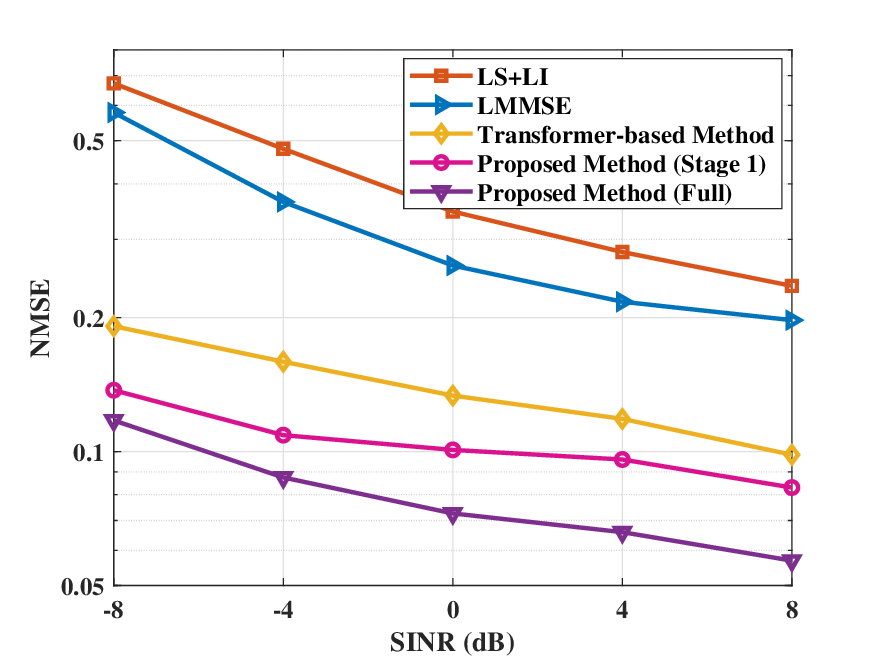}
			\caption{\small{NMSE versus SINR (dB) for frequency-domain channel prediction task.}}\vspace{-18pt}
		\end{center}
	\end{figure}

	\begin{figure}[t]
		\begin{center}
			\includegraphics[scale=0.57]{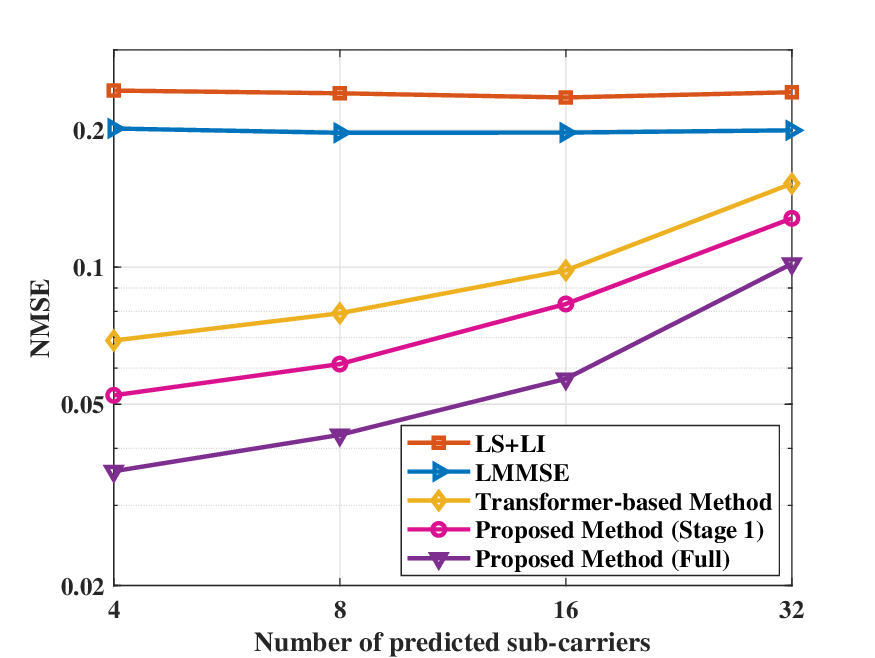}
			\caption{\small{NMSE versus percentage of the number of predicted sub-carriers ($N_f$) for frequency-domain channel prediction task.}}\vspace{-18pt}
		\end{center}
		\label{fig8}
	\end{figure}
	
	Fig. 10 shows the NMSE performance of frequency-domain channel prediction under different SINR levels, where the number of sub-carriers to be predicted $N_f$ is 16, and the percentage of downstream training samples $R_t$ is 50$\%$. It can be observed that the proposed method exhibits performance trends similar to those in the time-domain channel prediction task, consistently outperforming all baseline methods across the entire SINR range. In addition, the relative performance gains introduced by the proposed CSI refinement NN and the subsequent NPI suppression mechanism follow similar patterns to those in the time-domain task. Although frequency-domain channel prediction differs fundamentally from time-domain prediction in terms of prediction dimension, the proposed framework maintains stable and consistent advantages without requiring task-specific modifications. This indicates that the proposed method is capable of effectively capturing the intrinsic time–frequency correlation structure of wireless channels, thereby extracting clean, robust, and task-agnostic channel representations. As a result, the proposed framework exhibits strong generalization capability across different downstream tasks, validating its applicability as a unified wireless channel foundation model.
	
	Fig. 11 illustrates the NMSE performance with respect to the number of predicted sub-carriers ($N_f$), where the SINR is 8dB. In principle, reducing $N_f$ should lead to improved prediction accuracy, since the prediction task becomes less challenging. However, it can be observed that the NMSE of the conventional LS+LI and LMMSE methods remains almost unchanged. This is because the performance of these methods is fundamentally limited by the channel estimation error itself. The relatively large channel estimation error results in noisy and corrupted channel representations, which restrict the capability of the downstream prediction model. Consequently, even when predicting only a small number of sub-carriers, the prediction accuracy cannot be effectively improved.
	
	In contrast, according to Fig. 11, the proposed method benefits significantly from reducing $N_f$. By incorporating CSI refinement NN and NPI suppression mechanism, the proposed method produces cleaner and more structured channel representations that better preserve the underlying frequency-domain correlation. As the prediction range shrinks, the downstream model can effectively exploit frequency correlations, leading to more accurate predictions and larger performance gains. This behavior indicates that the proposed framework successfully alleviates the performance bottleneck of channel foundation model imposed by imperfect channel estimation.
	
	\subsection{Simulation Results - Outdoor Localization}
	
	\begin{figure}[t]
		\begin{center}
			\includegraphics[scale=0.57]{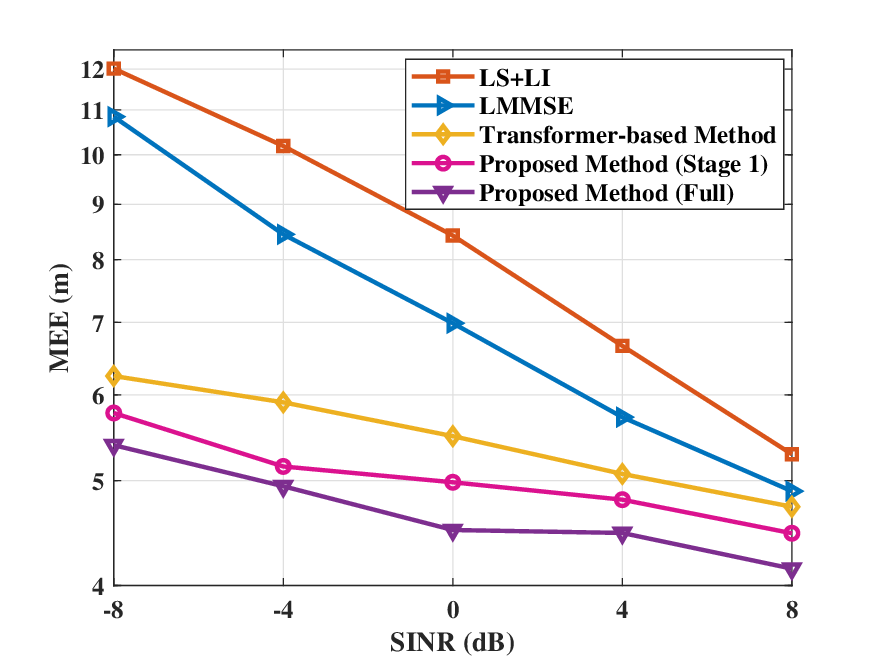}
			\caption{\small{MEE versus SINR (dB) for outdoor localization task.}}\vspace{-10pt}
		\end{center}
	\end{figure}

	Fig. 12 shows the MEE performance of outdoor localization with respect to SINR under the scenario of \emph{Oudoor 1}, and the percentage of downstream training samples $R_t$ is 50$\%$. As shown in Fig. 12, the localization error of all considered methods decreases with increasing SINR, since higher SINR leads to more accurate channel estimation and thus more informative channel representations for localization. However, the conventional LS+LI and LMMSE methods experience severe performance degradation in the low-SINR regime. This is because the localization task is highly sensitive to channel estimation errors, and the noisy and distorted CSI produced by traditional estimators significantly degrades the spatial features required for accurate position inference. Compared with traditional methods, the transformer-based approach is able to reduce the channel estimation error by exploiting data-driven feature extraction, thus improving the localization accuracy significantly under low-SINR conditions. Nevertheless, the proposed method still substantially outperforms the transformer-based approach, exhibiting an SINR gain of approximately 8 dB. By jointly performing CSI refinement and NPI suppression, the proposed method is able to preserve cleaner channel representations that are more robust and informative for localization, thereby fully demonstrating the superiority of the proposed framework. 
	
	\begin{figure}[t]
		\begin{center}
			\includegraphics[scale=0.57]{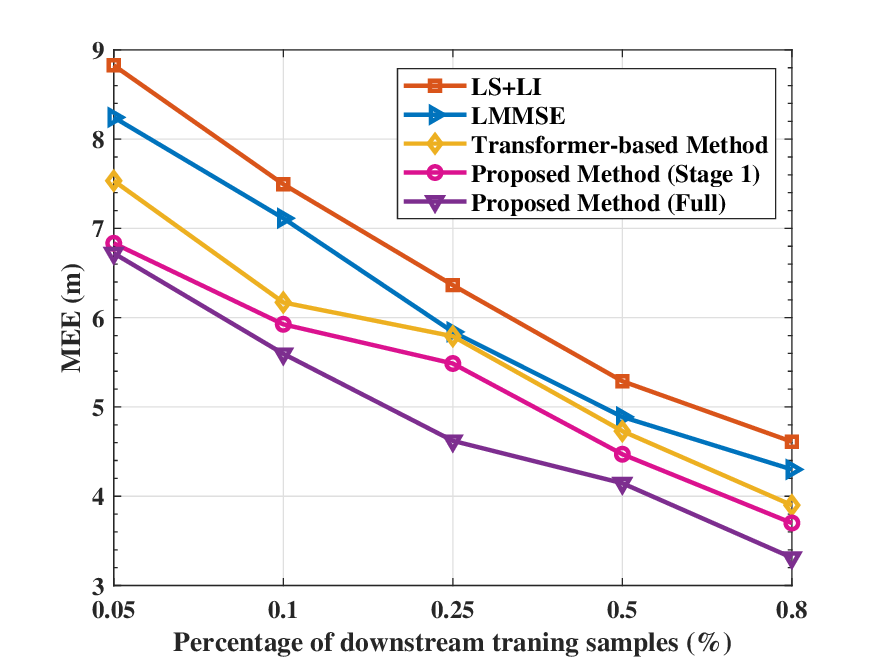}
			\caption{\small{MEE versus percentage of downstream training samples ($R_t$) for outdoor localization task.}}\vspace{-10pt}
		\end{center}
	\end{figure}

		\begin{figure}[b]
	\begin{center}
		\includegraphics[scale=0.53]{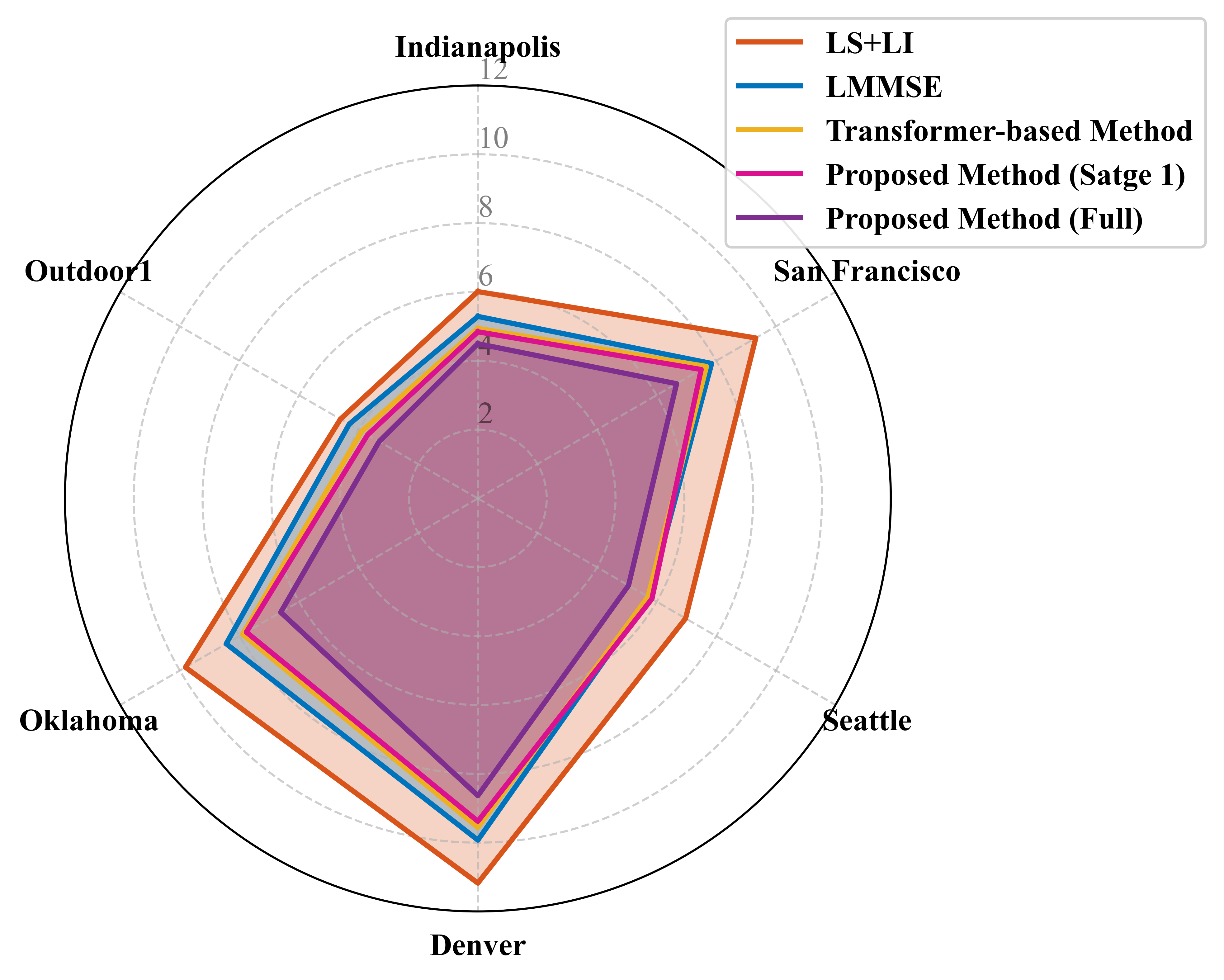}
		\caption{\small{Outdoor localization errors of various scenarios.}}
	\end{center}
\end{figure}

	Fig. 13 investigates the impact of percentage of downstream training samples ($R_t$) on the localization error, where SINR is 8dB and the scenario is \emph{Outdoor 1}. As is shown in Fig. 13, the performance curves of outdoor localization task are steeper than those of the time-domain channel prediction task in Fig. 9, indicating that localization is inherently more data-dependent. This can be attributed to the fact that localization requires high-level semantic inference from channel representations, rather than merely estimating low-level physical parameters, making it more sensitive to the amount of labeled training data. Despite the challenge, the proposed method outperforms all the other approaches across the entire range of $R_t$, demonstrating its strong cross-task generalization capability. This indicates that the unified channel representations extracted by the proposed framework are not dedicated to a specific downstream task, but are able to capture intrinsic and task-independent channel characteristics that can be effectively reused across various tasks.

	The radar chart in Fig. 14 provides a comprehensive performance comparison in terms of localization error over various scenarios, where the SINR is 8dB, and the percentage of downstream training samples $R_t$ is 80$\%$. As observed from Fig. 14, although the absolute localization errors vary across different scenarios due to environmental diversity, the relative performance ordering among the considered methods remains consistent. In particular, the proposed method achieves the lowest localization error in all considered scenarios, demonstrating the strong environmental generalization capability and robustness of the proposed method.

	\section{Concluding Remarks}
	
	In this paper, we proposed a wireless channel foundation model with noise-plus-interference suppression structure, which consists of a NPI estimation and suppression component, a CSI refinement/completion NN, and a SINR estimator. The proposed model works following a Filter-and-Attend paradigm. By jointly integrating channel estimation, denoising, and representation learning within a unified framework, the proposed model is able to extract accurate, robust, and task-independent channel representations under imperfect environments with limited pilot overhead. Extensive simulation results demonstrate that the proposed approach consistently outperforms existing methods across a variety of communication and sensing downstream tasks and diverse simulation settings.

	\ifCLASSOPTIONcaptionsoff
	\newpage
	\fi

\end{document}